\documentclass{article}
\usepackage{frascatiphys}

\def\vev#1{\left\langle #1\right\rangle}
\def\Im{\mathop{\mbox{Im}}}
\def\Re{\mathop{\mbox{Re}}}

\def\eoe{$\varepsilon'/\varepsilon$~}
\def\eps{$\varepsilon$~}
\def\be{\begin{equation}}
\def\ee{\end{equation}}
\def\bea{\begin{eqnarray}}
\def\eea{\end{eqnarray}}
\def\eq#1{eq.~\ref{#1}}

\def\Journal#1#2#3#4{{\rm #1} {\bf #2}, #3 (#4)}
\def\etal{{\it et al. }}
\def\NPB{Nucl. Phys. B}
\def\PLB{Phys. Lett. B}
\def\PRD{Phys. Rev. D}
\def\PRL{Phys. Rev. Lett.}

\begin{document}

\title{THEORY STATUS OF \eoe
}
\author{Stefano Bertolini \\
{\em INFN and SISSA, Via Beirut 4, I-34013 Trieste, Italy}}
\maketitle
\baselineskip=11.6pt
\begin{abstract}
I shortly review the present status of the theoretical
calculations of \eoe
and the comparison with the present experimental results. 
I discuss the role of higher order chiral
corrections and in general of non-factorizable contributions for
the explanation of the $\Delta I = 1/2$ selection rule 
and direct CP violation in kaon decays. 
Still lacking satisfactory lattice calculations,
analytic methods and phenomenological approaches are helpful in
understanding correlations among theoretical effects and
experimental data. Substantial progress from lattice QCD
is expected in the coming years.
\end{abstract}
\baselineskip=14pt
\section{Introduction}
The results obtained in the last few years by the NA48~\cite{NA48}
and the KTeV~\cite{KTeV} collaborations
have marked a great experimental achievement,
establishing some 35 years after the discovery of CP violation
in the neutral kaon system~\cite{Christenson}
the existence of a much smaller violation acting directly in the
decays:
\be
\label{eoenew}
\Re(\varepsilon'/\varepsilon) =
\left\{
\begin{array}{ll}
(15.3 \pm 2.6)\times 10^{-4} & {\rm (NA48)} \\
(20.7 \pm 2.8)\times 10^{-4} & {\rm (KTeV)} .
\end{array}
\right.
\ee
The average of these results with the previous measurements by
the NA31 collaboration at CERN
and by the E731 experiment at Fermilab
gives
\be
\label{eoewa}
\Re(\varepsilon'/\varepsilon) = (17.2 \pm 1.8)\times 10^{-4} .
\ee
While the Standard Model (SM) of strong and electroweak interactions
provides an economical and elegant understanding
of indirect~($\varepsilon$) and direct~($\varepsilon'$)
CP violation in term of a single phase,
the detailed calculation of the size of these effects
implies mastering strong interactions at a scale
where perturbative methods break down. In addition,
direct CP violation in $K\to\pi\pi$ decays
arises from a detailed balance of
two competing sets of contributions,
which may hopelessly inflate the uncertainties
related to the relevant hadronic matrix elements in the final outcome.
All that makes predicting \eoe a complex and
challenging task~\cite{review}.

Just from the onset of the calculation the presence in the
definition of \eoe , written as
\be
\frac{\varepsilon'}{\varepsilon} = \frac{1}{\sqrt{2}} \left\{
\frac{\langle ( \pi \pi )_{I=2} | {\cal H}_W | {K_L} \rangle}
{\langle ( \pi \pi )_{I=0} | {\cal H}_W | {K_L} \rangle}
 -
\frac{\langle ( \pi \pi )_{I=2} | {\cal H}_W | {K_S} \rangle}
{\langle ( \pi \pi )_{I=0} | {\cal H}_W | {K_S} \rangle} \right\}\ ,
\label{eoedef}
\ee
of given ratios of isospin
amplitudes warns us of a longstanding and still unsolved
theoretical ``problem'':
the explanation of the $\Delta I = 1/2$ selection rule.

The $\Delta I = 1/2$ selection rule in $K\to\pi\pi$ decays is known since
45 years~\cite{Pais-Gell-Mann} and it states the experimental
evidence that kaons
are 400 times more likely to decay in the $I=0$ two-pion state
than in the $I=2$ component ($\omega \equiv A_2/A_0 \simeq 1/22$).
This rule is not justified by any
symmetry argument and, although it is common understanding
that its explanation must be rooted in the dynamics of strong interactions,
there is up to date no derivation of this effect from first principle QCD.

Given the possibility that common systematic uncertainties may
a-priori affect the calculation of \eoe and the $\Delta I = 1/2$ rule
(see for instance the present difficulties in calculating on the lattice
the ``penguin contractions'' for CP violating as well as for
CP conserving amplitudes~\cite{Romapost})
a convincing calculation of \eoe must involve at the same time
a reliable explanation of the $\Delta I = 1/2$ selection rule.
Both observables indicate the need of large corrections
to factorization in the evaluation of the four-quark hadronic
transitions. Among these corrections
Final State Interactions (FSI) play a substantial role.
However, FSI alone are {\em not} enough to account
for the large ratio of the $I=0$ over $I=2$ amplitudes.
Other sources of large non-factorizable corrections are
therefore needed for the CP conserving 
amplitudes~\cite{review,instanton}, which
might affect the determination of \eoe as well.
As a consequence, a self-contained calculation of \eoe should also address
the determination of the $K\to\pi\pi$ rates.

\section{OPE: an ``effective'' approach}

The Operator Product Expansion (OPE) provides us with a very
effective way to address the calculation of hadronic transitions
in gauge theories. The integration of the ``heavy'' gauge  and
matter fields allows us to write the relevant amplitudes in terms
of the hadronic matrix elements of effective quark operators and
of the corresponding Wilson coefficients (at a scale $\mu$), which
encode the information about those dynamical degrees of freedom
which are heavier than the chosen renormalization scale. According to the
SM flavor structure the $\Delta S = 1$ transitions are effectively
described by \be {\cal H}_{\Delta S = 1} = \frac{G_{\rm
F}}{\sqrt{2}} V_{ud}\,V^*_{us} \sum_i \Bigl[{z_i}(\mu) + {\tau}\
{y_i}(\mu) \Bigr] {Q_i} (\mu) \ . \label{Leff} \ee The entries
$V_{ij}$ of the $3\times 3$ Cabibbo-Kobayashi-Maskawa (CKM) matrix
describe the flavour mixing in the SM and ${\tau} = -
{V_{td}V_{ts}^{*}}/V_{ud}V_{us}^{*}$. For $\mu < m_c$ ($q=u,d,s$),
the relevant quark operators are:
\be
\begin{array}{lcl}
\left.
\begin{array}{lcl}
{Q_{1}} & = & \left( \overline{s}_{\alpha} u_{\beta}  \right)_{\rm V-A}
            \left( \overline{u}_{\beta}  d_{\alpha} \right)_{\rm V-A}
\\[1ex]
{Q_{2}} & = & \left( \overline{s} u \right)_{\rm V-A}
            \left( \overline{u} d \right)_{\rm V-A}
\end{array}
\right\} &&\hspace{-1.6em} \mbox{Current-Current} \\[4ex]
\left.
\begin{array}{lcl}
{Q_{3,5}} & = & \left( \overline{s} d \right)_{\rm V-A}
   \sum_{q} \left( \overline{q} q \right)_{\rm V\mp A}
\\[1ex]
{Q_{4,6}} & = & \left( \overline{s}_{\alpha} d_{\beta}  \right)_{\rm V-A}
   \sum_{q} ( \overline{q}_{\beta}  q_{\alpha} )_{\rm V\mp A}
\end{array}
\right\} &&\hspace{-1.6em} \mbox{Gluon ``penguins''} \\[4ex]
\left.
\begin{array}{lcl}
{Q_{7,9}} & = & \frac{3}{2} \left( \overline{s} d \right)_{\rm V-A}
         \sum_{q} \hat{e}_q \left( \overline{q} q \right)_{\rm V\pm A}
\\[1ex]
{Q_{8,10}} & = & \frac{3}{2} \left( \overline{s}_{\alpha}
                                                 d_{\beta} \right)_{\rm V-A}
     \sum_{q} \hat{e}_q ( \overline{q}_{\beta}  q_{\alpha})_{\rm V\pm A}
\end{array}
\right\}  && \hspace{-1.6em} \mbox{Electroweak ``penguins''}
\end{array}
\label{quarkeff}
\ee
Current-current operators are induced by tree-level W-exchange whereas
the so-called penguin (and ``box'') diagrams are generated via an
electroweak loop.
Only the latter ``feel'' all three quark families via the virtual quark
exchange and are therefore sensitive to the weak CP phase.
Current-current operators control instead the CP conserving
transitions. This fact suggests already that the connection
between \eoe and the
$\Delta I = 1/2$ rule is by no means a straightforward one.

Using the effective $\Delta S=1$ quark Hamiltonian we can write \eoe as
\be
\frac{{\varepsilon'}}{\varepsilon} =
e^{i \phi} \frac{G_{\rm F} \omega}{2|\epsilon|\Re{A_0}} \:
{\mbox{Im}\, \lambda_t} \: \:
 \left[ {\Pi_0} - \frac{1}{\omega} \: {\Pi_2} \right]
\label{main}
\ee
where
\be
\begin{array}{lcl}
 {\Pi_0} & = & \frac{1}{{\cos\delta_0}}
\sum_i {y_i} \,
\Re\langle  Q_i  \rangle _0\ (1 - {\Omega_{\rm IB}})
\\[1ex]
 {\Pi_2} & = & \frac{1}{{\cos\delta_2}} \sum_i {y_i} \,
\Re\langle Q_i \rangle_2 \quad ,
\end{array}
\label{PI02}
\ee
and $\langle Q_i \rangle \equiv \langle \pi\pi | Q_i | K \rangle$.
The rescattering phases $\delta_{0,2}$ can be extracted from 
elastic $\pi$-$\pi$
scattering data\cite{FSIphases} and are such that $\cos\delta_0 \simeq 0.8$
and $\cos\delta_2 \simeq 1$. Given that the phase of $\varepsilon$
($\theta_\varepsilon$) is approximately $\pi/4$,
as well as the difference $\delta_0-\delta_2$, 
the $\varepsilon'/\varepsilon$ phase
$\phi = \frac{\pi}{2} + {\delta_2} - {\delta_0} - \theta_\varepsilon$
turns out to be consistent with zero.
While $G_{\rm F}$, $\omega$, $|\varepsilon|$ and $\Re A_0$ are precisely
determined by experimental data, the first source of uncertainty that
we encounter in \eq{main} is the value of
{$\Im \lambda_t \equiv \Im (V_{ts}^*V_{td})$},
the combination of CKM elements
which measures CP violation in $\Delta S = 1$ transitions.
The determination of $\Im \lambda_t$ depends on B-physics
constraints and on $\varepsilon$~\cite{Ciuchinietal}.
In turn, the fit of $\varepsilon$ depends on the
theoretical determination of $B_K$, the $\bar K^0-K^0$ hadronic parameter,
which should be self-consistently determined within every analysis.
The theoretical uncertainty on $B_K$ was in the past
the main component of the final uncertainty on $\Im \lambda_t$.
The improved determination of the unitarity triangle
coming from B-factories and hadronic colliders~\cite{CKM}
has weakened and will eventually lift the dependence of $\Im \lambda_t$
on $B_K$, allowing for an experimental measurement of the latter from
\eps .
Within kaon physics, the decay {$K_L\to\pi^0\nu\bar\nu$} gives the cleanest
``theoretical''
determination of $\Im\lambda_t$, albeit representing a great experimental
challenge. At present, a typical range of values for $\Im\lambda_t$
is $(0.94 - 1.60) \times 10^{-4}$~\cite{BurasLP01}.

We come now to the quantities in the square brackets.
While the calculation of the Wilson coefficients is well under control,
thanks primarily to the work done in the early nineties
by the Munich \cite{MunichNLO}
and Rome \cite{RomaNLO} groups, the evaluation of the ``long-distance''
factors in \eq{PI02} is the crucial issue for the ongoing calculations.
The isospin breaking (IB)
parameter $\Omega_{\rm IB}$, gives at the leading-order (LO)
in the chiral expansion a $positive$ correction to the $A_2$
amplitude (proportional to $A_0$ via the $\pi^0-\eta$ mixing)
of about 0.13~\cite{LOpieta}.
At the next-to-leading order (NLO)
the full inclusion of the $\pi^0-\eta-\eta'$
mixing lift the value of $\Omega_{\rm IB}$ to $0.16\pm 0.03$~\cite{NLOpieta}.
On the other hand, the complete NLO calculation
of IB effects beyond the $\pi^0-\eta-\eta'$ mixing
(of strong and electromagnetic origin, among
which the presence of $\Delta I = 5/2$ transitions)
involves a number of unknown NLO chiral couplings and is
presently quite uncertain.
Dimensional estimates show that IB effects
may be large and affect \eoe sizeably in both directions
~\cite{GardnerValencia}.
Although a partial cancellation
of the indirect ($\Delta\omega$) and direct $\Delta\Omega_{\rm IB}$
NLO isospin breaking corrections in \eq{main}
may reduce their final numerical impact on \eoe,
we must await for further analyses
in order to confidently assess their relevance.
At present one may use
$\Omega_{\rm IB}= 0.10\pm 0.20$~\cite{omegaNLOstr,omegaNLOmodel,omegaNLOem}
as a conservative estimate of the IB effects.

The final basic ingredient for the calculation of \eoe
is the evaluation of the $K\to\pi\pi$ hadronic matrix elements
of the quark operators in \eq{quarkeff}.
A simple albeit naive approach to the problem is
the Vacuum Saturation Approximation (VSA),
which is based on two drastic assumptions:
the factorization of the four quark operators
in products of currents and densities and the saturation
of the intermediate states by the vacuum state.
As an example:
\bea
\langle \pi^+ \pi^-|Q_6| K^0 \rangle & = &
 2\  \langle \pi^-|\overline{u}\gamma_5 d|0 \rangle
\langle \pi^+|\overline{s} u |K^0 \rangle
- 2\  \langle \pi^+ \pi^-|\overline{d} d|0 \rangle
\langle 0|\overline{s} \gamma_5 d |K^0 \rangle
\nonumber \\
& & +\ 2  \left[\langle 0|\overline{s} s|0 \rangle -
\langle 0|\overline{d}d|0 \rangle\right]
\langle \pi^+ \pi^-|\overline{s}\gamma_5 d |K^0 \rangle
\eea
The VSA does not exhibit
a consistent matching of the renormalization scale and scheme dependences
of the Wilson coefficients
and it carries potentially
large systematic uncertainties~\cite{review}.
On the other hand it provides useful insights on the main
features of the problem.
\begin{figure}
\vspace{5.5cm}
\includegraphics{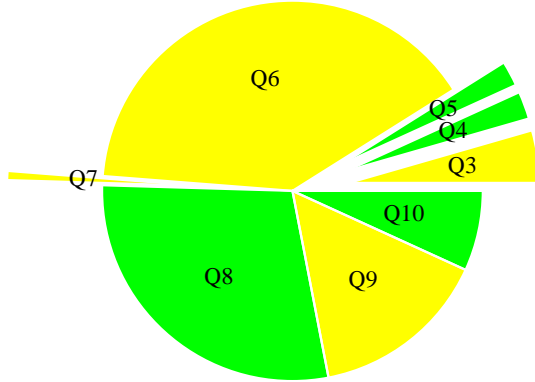}
\caption{Anatomy of \eoe in the Vacuum Saturation Approximation.
In light (dark) gray the positive (negative) contributions of the
effective four-quark operators are shown with proportional weight.
}\label{fig:pie}
\end{figure}
A pictorial summary of the relative weights of the contributions
of the various operators to \eoe, as obtained in the VSA,
is shown in Fig. \ref{fig:pie}.

As we have already mentioned, CP violation involves loop-induced
operators ($Q_3 - Q_{10}$). From Fig. \ref{fig:pie} one clearly notices
the potentially large cancellation
among the strong and electroweak sectors and the leading
role played by the gluonic penguin operator $Q_6$ and the
electroweak operator $Q_8$.
Tipical range of values for \eoe, obtained using the VSA, are
shown in Fig. \ref{fig:pre} together with the three most updated
predictions available before 1999 (when the first KTeV and NA48
results became known)~\cite{Munichpre,Romapre,ts98b}.
The fact that the cancellation among the strong and electroweak sectors
turns out to be quite effective (in the VSA) warns us about
the possibility that the uncertainties in the determination of the
relevant hadronic matrix elements may be largely amplified in the
calculation of \eoe. It is therefore important to asses carefully the
approximations related to the various parts of the calculations.
In particular, the analysis of the problem suggests that factorization
may be highly unreliable.

\section{Beyond Factorization}

\begin{figure}
\vspace{6.0cm}
\includegraphics{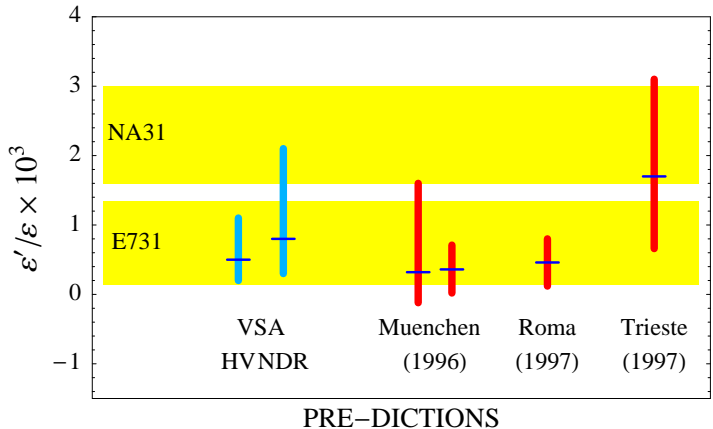}
\caption{
The 1-$\sigma$ results of the
NA31 and E731 Collaborations (early 90's)
are shown by the gray horizontal bands.
The old M\"unchen, Roma and Trieste theoretical predictions for \eoe are
depicted by the vertical bars with their central values.
For comparison, the VSA estimate is shown using two renormalization schemes.
}\label{fig:pre}
\end{figure}

The dark gray bars in Fig. \ref{fig:pre} depict the results
of three calculations of \eoe which are representative of approaches
that (in principle) allow us to go beyond naive factorization. They are based
from left to right on the large $N_c$
expansion~\cite{Munichpre,BBG},
on lattice regularization~\cite{Romapre,Roma},
and on phenomenological modelling of low-energy
QCD (the chiral quark model)~\cite{chiQM,PichEdR,ts98a,ts98b}.

\begin{figure}
\vspace{5.5cm}
\includegraphics{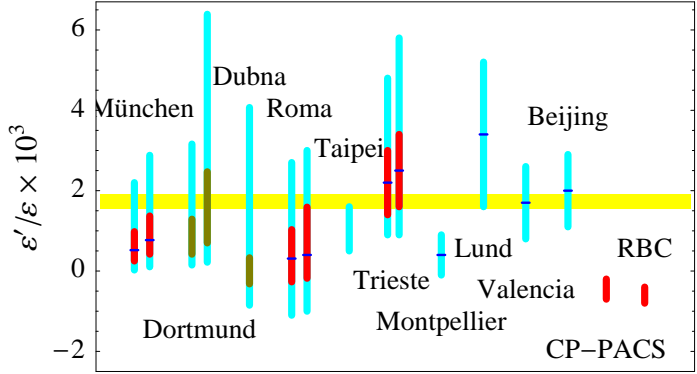}
\caption{
Recent theoretical calculations of \eoe are compared with
the combined 1-$\sigma$ average of the
NA31, E731, KTeV and NA48 results (\eoe = $17.2\pm 1.8\times 10^{-4}$),
depicted by the gray horizontal band.
}\label{fig:post}
\end{figure}

The experimental and theoretical scenarios have changed substantially
after the first KTeV data and the subsequent NA48 results.
Fig.~\ref{fig:post} shows the present experimental world average
for \eoe compared with the revised or new theoretical calculations
that appeared during the last year.
Without entering into the details of the results (for a short summary
see~\cite{radcor00}) they all represent attempts to incorporate
non-perturbative information into the calculation of the hadronic
matrix elements, whether their are based on the large $N_c$ expansion
(M\"unchen~\cite{Munichpost}, Dortmund~\cite{Dortmund},
Beijing~\cite{Beijing}, Taipei~\cite{Taipei}, Valencia~\cite{Valencia}),
phenomenological modelling of low-energy QCD
(Dubna~\cite{Dubna}, Trieste~\cite{ts98b,ts00b}, Lund~\cite{Lund}),
QCD Sum Rules (Montpellier~\cite{Montpellier}) or, finally, on lattice
regularization (Roma~\cite{Romapost}, CP-PACS~\cite{CP-PACS},
RBC~\cite{RBC}).

Overall most of the theoretical calculations
are consistent with a non-vanishing positive effect in the SM
(with the exception of the recent lattice
results on which I will comment shortly).

At a closer look however, if we focus our attention on
the central values,
many of the predictions
prefer the $10^{-4}$ regime, whereas only a few of them stand
above $10^{-3}$.
Is this just ``noise'' in the theoretical calculations?
Without entering the many details on which the estimates are based,
most of the aforementioned difference can be  explained in terms
of a single effect: the different size of the
hadronic matrix element of the gluonic penguin $Q_6$ as
obtained in the various approaches. In turn, this can be understood
in terms of sizeable higher order chiral contributions (NLO in
the $1/N_c$ expansion) to the $I=0$ amplitudes.

This effect was stigmatized well
before the latest experimental round
by the work of the Trieste group~\cite{ts98a,ts98b},
and appears clearly
in the comparison of the leading $1/N_c$ and lattice
results with the chiral quark model analysis in Fig.~\ref{fig:pre}.
The chiral quark model approach, together
with the fit of the CP conserving amplitudes which
normalizes phenomenologically the matching and the model parameters,
allows us to carry the calculation of the hadronic matrix elements
beyond the leading order in the chiral expansion (including the needed
local counterterms). Non-factorizable chiral contributions
(missing in the leading $1/N_c$ or lattice calculations)
were shown to produce a substantial enhancement
of the $I=0$ transitions thus lifting the expectation of \eoe
at the $10^{-3}$ level.

Since then a number of groups have attempted to improve
the calculation of $K\to\pi\pi$ matrix elements in a
model independent way.
Table \ref{tab:B6B8} presents a comparison of
different calculations of the relevant matrix elements.
Due to the leading role played by $Q_6$ and $Q_8$ we may write a
simplified version of \eq{main}, 
\be
\frac{\varepsilon'}{\varepsilon} \approx 13
\left(\frac{\Lambda_{\overline{MS}}^{(4)}}{340\ {\rm MeV}}\right)
\Im \lambda_t \left[\frac{110\ {\rm MeV}}{m_s\ (2\ {\rm
GeV})}\right]^2\ \left[{B_6} (1-\Omega_{\rm IB}) - 0.4
{B_8^{(2)}}\right]\ , 
\vspace*{-1ex}
\ee 
which although
``not be used for any serious analysis''~\cite{Munichpost} gives
an effective and practical way to test and compare different
calculations.

The B-factors $B_i \equiv  \vev{Q_i}/\vev{Q_i}_{\rm VSA}$
represent a convenient parametrization of the hadronic matrix
elements, albeit tricky, in that their values are in general scale
and renormalization-scheme dependent, 
and a spurious dependence on the quark masses is
introduced in the result whenever quark densities are involved. 
The latter is the case for the $Q_6$ and $Q_8$ penguins. 
As a consequence the VSA normalization
may vary from author to author thus introducing systematic
ambiguities.
By taking the VSA matrix elements at the scale $\mu=2$ GeV
we obtain~\cite{review}
\be
\begin{array}{lcl@{}}
\hspace*{-3em} &&
\vev{(\pi\pi)_{2} | Q_8 | K^0}_{\rm VSA} = \sqrt{6}\ f\ m_K^4\
  (m_s+m_d)^{-2} \simeq 1.1\ {\rm GeV}^3 \ ,   \\[1ex]
\hspace*{-3em} &&
\vev{\pi\pi| Q_6 | K^0}_{\rm VSA}/
  \vev{(\pi\pi)_{2} | Q_8 | K^0}_{\rm VSA}
  = -2\sqrt{2}\ (f_K-f_\pi)/f_\pi \simeq -0.63 \ ,
\end{array}
\label{VSAnorm}
\ee
where I have used
{$(m_s+m_d)(2\ {\rm GeV}) = 110\ {\rm MeV}$} and
the chiral value $f = 86$ MeV for the octet decay constant.

It is known that $B_6$ and $B_8^{(2)}$ are
perturbatively very weakly dependent on the
renormalization scale~\cite{review}. Therefore it makes sense
to compare the $B$'s obtained in different approaches, where
the matrix elements $\vev{Q_i}$ are computed at different scales.
The results for the
relevant penguin matrix elements coming from various approaches
are collected in Table~\ref{tab:B6B8}, paying
care to normalizing the data in a homogeneous way (as far as detailed
information on definitions and renormalization schemes was available).

As a guiding information, taking $\Im \lambda_t = 1.3\times 10^{-4}$,
the present experimental central value of \eoe is reproduced by
$B_6\ (1-\Omega_{\rm IB}) - 0.4\ B_8^{(2)} \approx 1 $.

\begin{table}[t]
\label{tab:B6B8}
\caption{Comparison of various calculations of penguin matrix elements.
The data marked by the star are rescaled by a factor $\sqrt{3/2}$, to
account for a different definition of the isospin matrix elements.}
\begin{tabular}{@{}lllll}
\hline
Method & {$B_{6}$} (NDR) & & {$B_{8}^{(2)}$} (NDR) & \\
\hline
Lattice (DWF, $K\to\pi$) & {$< 0.3$} & & $\sim 0.9$  & 
CP-PACS\cite{CP-PACS} \\
Lattice (DWF, $K\to\pi$) & $\sim 0.4$ & & $\sim 1$ & 
RBC\cite{RBC}\\
Lattice ($K\to\pi + \chi$PT)& $-$ & & $0.58\pm 0.06$ *  & 
APE\cite{APE} \\
Lattice ($K\to\pi + \chi$PT)& $-$ & & $0.56\pm 0.07$ * & 
SPQcdR\cite{SPQcdR} \\
Lattice ($K\to\pi\pi$) & $-$ & & $0.64\pm 0.07$ * & 
SPQcdR\cite{SPQcdR} \\
Large $N_c$+LMD ($\chi$-limit)  & $-$ & & $2.6 \pm 0.8$ * &  
Marseille\cite{PerisEdR} \\
Dispersive+data ($\chi$-limit)  & $-$ & & $2.5\pm 0.8$ * &  
Amherst\cite{CDGM} \\
Dispersive+data ($\chi$-limit)  & $-$ & & {$1.4\pm 0.6$} & 
Lund\cite{BGP} \\
Large $N_c$ + data  & $1.0\pm 0.3 $ & & $0.8\pm 0.2$ & 
Munich\cite{Munichpost} \\
NLO $1/N_c$ CHPT  & $1.5 \sim 1.7$ & & $0.4 \sim 0.7$ & 
Dortmund\cite{Dortmund} \\
NLO $1/N_c$ ENJL ($\chi$-limit) & $2.9\pm 0.5$ & & $1.5\pm 0.2$ &
Lund\cite{Lund} \\
NLO $\chi$QM + $\chi$PT  & $1.5\pm 0.4$ & & $0.84\pm 0.04$ &
Trieste\cite{ts98b} \\
Large $N_c$ + FSI & $1.55\pm 0.10$ & & $0.92\pm 0.03$ & 
Valencia\cite{Valencia} \\
\hline
\end{tabular}
\end{table}

The most important fact is the first evidence of a signal in lattice
calculations of $\vev{\pi|Q_6|K}$, obtained by the CP-PACS~\cite{CP-PACS}
and RBC~\cite{RBC} collaborations. 
Both groups use the Domain Wall Fermion approach
which allows to control the chiral symmetry on the lattice as a volume
effect in a fifth dimension. This approach softens in principle
the problem of large power subtractions which affects the lattice
extraction of $I=0$ amplitudes (penguin contractions).
Still only the $\vev{\pi|Q_i|K}$ transition is computed
on the lattice and LO chiral perturbation theory is used to extrapolate
it to the physical amplitude. The two groups obtain comparable values
of the $Q_6$ and $Q_8$ matrix elements leading both to a negative \eoe
(and do not agree on the CP conserving $I=0$ amplitude).
On the other hand the calculations are at an early stage and do not
include higher order chiral dynamics which may be responsible
for the enhancement of $I=0$ amplitudes (as large $N_c$
approaches beyond LO and the Chiral Quark Model strongly suggest).
The SPQcdR collaboration has reported a result for the $Q_8$
matrix element from direct calculation of the $K\to\pi\pi$
amplitude on the lattice. This result agrees with previous
lattice data, albeit it does not yet include the chiral
corrections relevant to quenching and to the extrapolation to the physical
pion mass~\cite{SPQcdR}.

Among the analytic approaches
important results have been obtained using data on spectral functions
in connection with QCD sum rules and dispersive relations in the attempt
to obtain model independent information on the relevant matrix elements.
These approaches have produced as of today calculations of $Q_8$
(in the chiral limit)
which are subtantially larger than the factorization (and lattice)
results. While there is still disagreement among
the different analysis,
we must await the calculation of the $Q_6$ matrix element and a
quantitative assessment of chiral breaking effects before drawing 
conclusions on these as well as lattice results.

Calculations which sofar have allowed for the determination of all
relevant parameters,
based on chiral perturbation theory and/or models of low-energy
QCD, have shown the crucial role of higher order non-factorizable corrections
in the enhancement of the $I=0$ matrix
elements \cite{ts98b,Dortmund,Valencia,Lund}. Chiral loop corrections
drive the final value of \eoe in the ballpark of the present data.
However the calculation of higher order chiral effects cannot be
fully accomplished in a model independent way due to the many unknown
NLO local couplings. In the chiral quark model approach all needed
local interactions are computed in terms of quark masses, meson decay constants
and a few non-perturbative parameters as quark and gluon condensates.
The latter are determined self-consistently in a phenomenological way
via the fit of the CP conserving $K\to\pi\pi$ amplitudes, thus encoding
the $\Delta I =1/2$ rule in the calculation~\cite{ts98a,ts98b}. 
The analysis shows that
the role of local counterterms is subleading to the chiral logs when using
the Modified Minimal Subtraction (as opposed to the commonly used
Gasser-Leutwyler prescription). The phenomenological
fit is crucial in stabilizing the numerical prediction~\cite{ts98b}. 
The fact that
the model parameters (quark and gluon condesates, constituent quark mass)
turn out to be in the expected range, shows that the explicitly included
chiral (and $1/N$ gluon condensate) 
corrections represent the largest non-factorizable effect. 

Among higher order corrections FSI play a leading role.
As a matter of fact, one should in general expect an enhancement of
\eoe with respect to the naive VSA due to FSI.
As Fermi first argued~\cite{Fermi}, in potential scattering
the isospin $I=0$ two-body~states feel an
attractive interaction, of a sign opposite to that of the $I=2$
components thus affecting the size of the corresponding amplitudes.
This feature is at the root of the enhancement of the
$I=0$ amplitude over the $I=2$ one and of the corresponding enhancement
of \eoe beyond factorization.
An attempt to resum these effects in a model independent way
has been worked out by the authors of ref.~\cite{Valencia},
using a dispersive approach a la Omn\`es-Mushkelishvili~\cite{Omnes,Truong}.
Their analysis shows that resummation does not substantially modify
the one-loop perturbative result and, as it appears from Table \ref{tab:B6B8},
a 50\% enhancement of the gluonic penguin matrix element is found over
the factorized result.
However,
the calculation suffers from a sistematic uncertainty due to
the indetermination of the off-shell amplitude 
which is identified with the large $N_c$ result~\cite{FSIcritics}.
Even when the authors in the most recent work match the dispersive
resummation with the on-shell perturbative one-loop calculation, 
thus including $1/N_c$ effects, again
a systematic uncertainty remains in the unknown polinomial
parts of the local chiral counterterms.
Therefore, a model-independent complete calculation of chiral loops
for $K\to\pi\pi$ is still missing.

Finally, it has been recently emphasized~\cite{dim8}
that cut-off based approaches should
pay attention to higher-dimension
operators which become relevant for matching scales below 2 GeV
and may represent one of the largest sources of uncertainty in present
calculations. The results of refs. \cite{CDGM,BGP} include these effects.
The calculations based on dimensional regularization
may be safe if phenomenological input is used in order to encode
in the relevant hadronic matrix elements the physics at all scales
(this is done in the Trieste approach).

In summary, while model dependent calculations suggest no conflict
between theory and experiment for \eoe, a precise and "pristine" prediction
of the observable is still quite ahead of us.

\section{Outlook and Conclusions}

Higher-order chiral corrections are taking the stage of $K\to\pi\pi$ physics.
They are needed in order to asses the size of crucial parameters
(as $\Omega_{\rm IB}$) and the effect of non-factorizable contributions
in the penguin matrix elements.

Lattice, as a regularization of QCD, is {\em the} first-principle
approach to the problem. However,
lattice calculations still heavily depend
on chiral perturbation theory \cite{Golterman}.
Presently, very promising developments are being undertaken
to circumvemt the technical and conceptual shortcomings related
to the calculation of weak matrix elements~\cite{SPQcdR,Sachrajda}.
Among those are the Domain Wall Fermion
approach~\cite{DWF} which allows us to decouple the chiral symmetry
from the continuum limit, and the very interesting observation that
the Maiani-Testa theorem~\cite{MaianiTesta} can be overcomed using the fact that
lattice calculations are performed in finite volume~\cite{LellouchLuscher},
thus allowing for the direct calculation of the physical $K\to\pi\pi$ amplitude
on the lattice.
All these developments need a tremendous effort in machine power
and in devising faster algorithms. Preliminary results
for lattice calculations of both \eoe
and the $\Delta I = 1/2$ selection rule
are already available and others are currently under way \cite{SPQcdR}.

In the meantime analytical and semi-phenomenological approaches
have been crucially helpful in driving the attention of the
community on some systematic short-comings of "first-principle"
calculations. The amount of theoretical work triggered by the
NA48 and KTeV data promises rewarding and perhaps exciting results
in the forthcoming years.



\end{document}